\documentclass{elsart}
\def\gg{\gamma\gamma}
\def\H{\rm h}

\def\g{\mbox{g}}
\def\q{\rm q}
\def\s{\rm s}
\def\h{\rm h}
\def\ccbar{\overline{\mbox c}\mbox{c}}
\def\bbbar{\overline{\mbox b}\mbox{b}}
\def\qqbar{\overline{\mbox q}\mbox{q}}
\def\ccbarg{\overline{\mbox c}\mbox{cg}}
\def\bbbarg{\overline{\mbox b}\mbox{bg}}
\def\BR{\rm BR}

\def\pz{\phantom{0}}

\usepackage{epsfig}
\begin{document}
\runauthor{Cicero, Caesar and Vergil}
\begin{frontmatter}
\title{Light Higgs Production at a Photon 
Collider\thanksref{X}}
\author[Stefan]{Stefan S\"oldner-Rembold\thanksref{Someone}}
\author[Georgi]{Georgi Jikia}
\address[Stefan]{CERN, CH-1211 Geneva 23, Switzerland}
\address[Georgi]{Albert--Ludwigs--Universit\"at Freiburg, 
                 Hermann--Herder--Str.\ 3, 
                 D--79104 Freiburg, Germany}
\thanks[X]{Talk given at the International Workshop on 
High Energy Photon Colliders, June 14 - 17, 2000,
DESY Hamburg, Germany }
\thanks[Someone]{Heisenberg Fellow}
\begin{abstract}
We present a preliminary study 
of the production of a light Higgs boson with a mass
between 120 and 160 GeV
in photon-photon collisions at a Compton collider. The event
generator for the backgrounds to a Higgs signal due to $\bbbar$ and 
$\ccbar$ heavy quark pair production in polarized $\gamma\gamma$
collisions is based on a complete next-to-leading order (NLO) perturbative
QCD calculation. For $J_z=0$ the large double-logarithmic corrections
up to four loops are also included.  It is shown that the
two-photon width of the Higgs boson 
can be measured with high statistical accuracy of about $2-10\%$ for
integrated $\gg$ luminosity in the hard part of the spectrum of
$43$~fb$^{-1}$. From this result the total Higgs boson width can be derived
in a model independent way.
\end{abstract}
\end{frontmatter}

\section{Introduction}
\typeout{SET RUN AUTHOR to \@runauthor}
The experimental discovery of the Higgs boson is crucial for 
understanding the mechanism of electroweak symmetry breaking. The
search for Higgs particles is one of the main goals for LEP2 and the
Tevatron and will be one of the major motivations for the future Large
Hadron Collider (LHC) and Linear e$^+$e$^-$ Collider (LC). Once the
Higgs boson is discovered, it will be of primary importance to
determine its tree-level and one-loop
induced couplings, spin, parity, CP-nature, and its
total width in a model independent way. 
In this respect, the $\gamma\gamma$ Compton 
Collider option of the LC offers a unique possibility to produce the
Higgs boson as an $s$-channel resonance decaying into $\bbbar$,
WW$^*$ or ZZ
and thereby to measure the two-photon Higgs width. This partial width
is of special interest, since it first appears at the one-loop level so
that all heavy charged particles which obtain their masses from
electroweak symmetry breaking contribute to the loop. Moreover, the
contributions of very heavy particles do not decouple. In addition,
combined measurements of $\Gamma(\h\to\gamma\gamma)$ and
$\BR(\h\to\gamma\gamma)$ at the LC provide a model independent
measurement of the total Higgs width~\cite{snow96}.

The lower bound on $m_{\h}$ from
direct searches at LEP is $113.5$~GeV at $95\%$~CL~\cite{pik}. 
A recent global analysis of precision electroweak data~\cite{osaka}
suggests that the Higgs boson is light, yielding 
$m_{\rm h}=62^{+53}_{-30}$~GeV. 
This fact is in remarkable agreement with the well known upper bound of
$\sim 130$ GeV for the lightest Higgs boson mass in the minimal
version of supersymmetric theories, the Minimal Supersymmetric
Standard Model (MSSM)~\cite{MSSM}. For this case of a light Higgs
boson the results of Monte Carlo simulations of Higgs
production in $\gamma\gamma$ collisions with final decay to $\bbbar$
quark pairs will be presented here. 
Similar studies as presented here but without detector simulation
can be found in~\cite{melles} but without detector simulation and
in~\cite{watanabe} but without taking into account higher order double
logarithmic corrections.
Since the current upper bound on the Higgs mass from radiative
corrections is $m_{\h}<170$~GeV at $95\%$~CL~\cite{osaka}, one can still 
hope to measure the two-photon Higgs width at the 300-500 GeV LC for heavier
Higgs masses by studying its production in $\gamma\gamma$
collisions and final decays into WW$^*$~\cite{WW*} or ZZ~\cite{ZZ}
states.

The accuracy of the $\Gamma(\h\to\gamma\gamma)$ measurements to be reached
can be inferred from the results of the studies of the coupling of the
lightest SUSY Higgs boson to two photons in the decoupling regime
\cite{decoupling}. It was shown that in the decoupling limit, where all
other Higgs bosons are very heavy and no supersymmetric particle has
been discovered at LHC or LC, chargino and top squark loops can
generate a sizable difference between the standard and the SUSY
two-photon Higgs couplings. Typical deviations are at the few percent
level. Top squarks heavier than 250 GeV can induce deviations even
larger than $\sim 10\%$ if their couplings to the Higgs boson are large.

\section{Signal and background}

The cross-section of resonant Higgs production at a $\gamma\gamma$
Collider is proportional to the product
\begin{equation}
\sigma(\gamma\gamma\to \h^0\to X) =
 z \frac{dL_{\gamma\gamma}}{dz}
\frac{4\pi^2}{M_{\h^0}^3} \Gamma(\h^0\to\gamma\gamma)
\cdot \BR(\h^0\to X)(1+\lambda_1\lambda_2)
\end{equation}
Here the effective photon-photon luminosity $L_{\gamma\gamma}$ is
introduced (see the next section). $\lambda_{1,2}$ are mean high
energy photon helicities.

The Standard Model (SM) Higgs branching ratios and the Higgs total width are
calculated with the help of the program HDECAY~\cite{HDECAY}. The
program includes the full massive NLO corrections for $\h\to \qqbar$
decays close to the thresholds as well as the massless ${\mathcal
O}(\alpha_{\rm s}^3)$ corrections far above the thresholds. For the Higgs
signal only two-particle final states are generated, since the Parton
Shower (PS) algorithm of JETSET is used to simulate three and higher
particle final states.

The main background to $\h$ production is the continuum
production of $\bbbar$ and $\ccbar$ pairs. In this respect, the
availability of high degree of photon beams circular polarization is
crucial, since for the equal photon helicities $(\pm\pm)$ producing
spin-zero resonant states, the $\gamma\gamma\to \qqbar$ QED Born
cross-section is suppressed by the factor $m_{\q}^2/s$
\cite{BBC}:
\begin{equation}
\frac{d\sigma^{\rm Born}(J_z=0)}{dt} =
\frac{12 \pi \alpha^2 Q_{\q}^4}{s^2}
\frac{m_{\q}^2 s^2(s-2 m_{\q}^2)}{t_1^2 u_1^2}
\end{equation}
and
\begin{equation}
\frac{d\sigma^{\rm Born}(J_z=\pm2)}{dt} = 
\frac{12 \pi \alpha^2 Q_{\q}^4}{s^2}
\frac{(t_1 u_1 - m_{\q}^2 s)(u_1^2+t_1^2+2 m_{\q}^2 s)}
{t_1^2 u_1^2}.
\end{equation}
Here $m_{\q}$ is the quark mass, $Q_{\q}$ its charge, and $t_1 = t-m_{\q}^2$,
$u_1 = u-m_{\q}^2$.

Virtual one-loop QCD corrections for $J_z=0$ are found to be
especially large due to the double-logarithmic enhancement factor, so
that the corrections are comparable or even larger than the Born
contribution for the two-jet final topologies~\cite{JikiaTkabladze}. 
For small values of the cutoff $y_{\rm cut}$
separating two and three-jet events, the two-jet cross-section calculated
to order $\alpha_{\s}$ becomes negative in the central region.
Recently leading double-logarithmic QCD corrections for $J_z=0$ have been
resummed to all orders~\cite{Sudakov,melles}. 
Taking into account non-Sudakov form
factors to higher orders makes the cross-section well defined and
positive definite in all regions of the phase space.

The simulation program includes exact one-loop QCD corrections to
heavy quark production in $\gamma\gamma$ collisions~\cite{JikiaTkabladze} 
and the non-Sudakov form factor in the
double-logarithmic approximation through four loops~\cite{Sudakov}:
\begin{eqnarray}
\lefteqn{\frac{\sigma^{\rm DL}_{\rm virt}}{\sigma_{\rm Born}} \sim 
1 + 6 {\mathcal F} 
+
\frac{1}{6} \left(56 +2 \frac{C_A}{C_F} \right) {\mathcal F}^2 + }\\ &&  
\frac{1}{90} \left( 94 +90 \frac{C_A}{C_F} +2 \frac{C_A^2}{C_F^2} \right)
{\mathcal F}^3 
+\frac{1}{2520} \left( 418 + 140 \frac{C_A}{C_F} + 238 \frac{C_A^2}{C_F^2}
+3 \frac{C_A^3}{C_F^3} \right) {\mathcal F}^4 \nonumber
\end{eqnarray}
where ${\mathcal F} = - C_F \frac{\alpha_{\s}}{4 \pi} \log^2 \frac{m_{\q}^2}{s}$
is the one-loop hard form factor. Since it is a non-trivial task to
write down an event generator including both NLO corrections and
the Parton Shower algorithm, we do not use any Parton Shower for
background $\bbbar$ and $\ccbar$ production. So the experimental value
of the $y_{\rm cut}$ parameter should not be chosen too small, otherwise 
resummed Sudakov corrections are needed. Two-parton ($\bbbar$,
$\ccbar$, and $\bbbarg$, $\ccbarg$ with $y_{\rm cut}<0.01$) and 
three-parton ($\bbbarg$, $\ccbarg$ with $y_{\rm cut}>0.01$) final states
are generated separately and the JETSET~\cite{PYTHIA} string
fragmentation algorithm is applied afterwards. The event generator
both for the Higgs signal and heavy quark background is
implemented using the programs BASES/SPRING~\cite{BASES}.

\section{\boldmath $\gamma\gamma$ \unboldmath luminosity}

The original polarized photon energy spectra~\cite{luminosity} are
used assuming $100\%$~laser and $85\%$~electron beam polarizations 
with $2\lambda_e^{1,2}\lambda_\gamma^{1,2}=-0.85$. The Parameter
$x=\frac{4E_e\omega_0}{m_e^2}$ is taken to be 4.8.  Assuming
that the Higgs boson will already have been discovered at LEP, the LHC and/or 
the LC and that
its mass will be known, we tune the ee collision energy $\sqrt{s}_{\rm ee}$
to be $\sqrt{s}_{\rm ee}=m_{\h}/0.8$
so that the Higgs mass corresponds to the
peak of the photon-photon luminosity spectrum
$z\frac{dL_{\gamma\gamma}}{dz}$, $z=0.8$, where 
$$z=\frac{W_{\gamma\gamma}}{2E_e}, \quad z_{\rm max}=\frac{x}{x+1}=0.83.$$
We assume a total integrated $\gamma\gamma$ luminosity of 
$L_{\gamma\gamma}(0<z<z_{\max}) = 150$~fb$^{-1}$ 
with idealized Compton spectra~\cite{luminosity}. 
Realistic simulations of the $\gamma\gamma$ luminosity~\cite{Telnov}
taking into account beamstrahlung, coherent pair creation and
interaction between charged particles show that idealized 
spectra~\cite{luminosity} will be strongly distorted in the low energy part of
the spectrum. However, in the hard part of the spectrum which is relevant
for our simulation, the idealized spectra represent a
very good approximation~\cite{Telnov}. The
luminosity in the hard part of the spectrum is
$L_{\gamma\gamma}(0.65<z<z_{\rm max}) = 43$~fb$^{-1}$ which 
corresponds to the integrated geometric  luminosity of e$^-$e$^-$
collisions of $L_{\rm ee} \approx 400$~fb$^{-1}$
\footnote{According to the 
present understanding~\protect\cite{highlumi1,highlumi2} 
$L_{\rm ee}$ at TESLA  in the
Higgs region can be about  $L_{{\rm e}^+{\rm e}^-}(500\ \mbox{GeV})$, 
where "nominal" $L_{{\rm e}^+{\rm e}^-}(500\ \mbox{GeV}) = 3\times
10^{34}$~cm$^{-2}$s$^{-1}$.}.

\section{Cross-sections}

In Table 1 the cross-sections for the
Higgs signal and for the background calculated without detector simulation
are given.
Both quark jets are assumed to satisfy a $|\cos\theta|<0.9$ cut. 
The resolved photon
contribution to the $\bbbar$ background is found to be negligible.
It is therefore not
included in the subsequent detector simulation analysis. 

\section{Results of the Monte Carlo simulation}

The Monte Carlo simulation of the fragmentation
is done with JETSET~\cite{PYTHIA}. Signal and
background are studied using the fast detector simulation program
SIMDET~\cite{SIMDET} for a typical TESLA detector. 
Jets are reconstructed using the Durham algorithm
with $y_{\rm cut}=0.02$. b tagging is not simulated and b tagging
efficiencies of $70\%$ for $\bbbar$ events and $3.5\%$ for $\ccbar$
events are used instead~\cite{Battaglia}. These efficiencies are based
on double-tagging of b jets to suppress $\ccbar$ events 
by a factor of 20 which is large enough to overcome the
enhancement factor of approximately 16 due to the larger c quark charge. 
The c rejection and the b tagging efficiency depend very much
on the inner radius of the tracking detector~\cite{hawkings}.
 
The jet multiplicities for a Higgs signal with $m_{\h}=120$~GeV
and for the heavy quark background at the detector level are
shown in Fig.~1. The three-jet rates
are comparable for signal and background. Applying a
jet multiplicity cut ($n_{\rm jet}=2$) will therefore not
improve significantly the signal to background ratio for the chosen
$y_{\rm cut}$ value. 
Another advantage of not applying a jet multiplicity cut is due
to the fact that no resummation of Sudakov logarithms was done
for the heavy quark background. If no jet multiplicity cut is used,
Sudakov logarithms are not present. 

The t-channel background processes are forward peaked 
and the NLO order corrections even increase this effect, whereas
the s-channel signal process has an isotropic angular distribution.
The following cuts are therefore used to suppress background:

\noindent (i) Events where quark jets are scattered at a small angle are
rejected by requiring for the thrust angle $|\cos\theta_{\rm T}|<0.7$
(see Fig.~2).

\noindent (ii) Since the Higgs boson is produced at the peak of
the photon-photon
luminosity spectrum almost at rest, a cut on the 
longitudinal momentum component
of the event divided by the ee centre-of-mass energy,
$|p_{\rm z}|/\sqrt{s}_{\rm ee}<0.1$, further reduces the background.

\noindent (iii) Most background events are produced at the lower energy tail
of the photon-photon luminosity distribution. A cut on the total visible
energy, $E_{\rm vis}/\sqrt{s}_{\rm ee}>0.6$, eliminates most
soft background events.

The invariant mass distributions for the combined $\bbbar (\g)$
and $\ccbar (\g)$ background, and for the Higgs signal 
are shown in Fig.~3 after applying these cuts.
The relative statistical error is calculated using
$$
\frac{\Delta\Gamma(\h\to\gg)}{\Gamma(\h\to\gg)}= 
\frac{\sqrt{N_{\rm obs}}}{N_{\rm obs}-N_{\rm BG}},
$$
where $N_{\rm obs}$ is the sum of the signal and background events
and $N_{\rm BG}$ the number of background events.
It lies in the range
$$\frac{\Delta[\Gamma(\h\to\gg)\BR(\h\to
\bbbar)]}{[\Gamma(\h\to\gg)\BR(\h\to\bbbar)]}\approx 2-10\%$$
in the Higgs mass range between 120 and 160~GeV.  

Systematic errors would include the modelling of the b-tagging
and the precise determination of the background shape. 
The background shape could be studied
without relying on the theoretical predictions
by running the collider at
an energy below the Higgs production threshold, since
the light Higgs boson is expected to be very narrow.

Assuming that at the e$^+$e$^-$ linear
collider (or at a high luminosity $\gg$ collider) 
the $\h\to \bbbar$ and $\h\to\gg$ branching
ratios can be measured with an accuracy of
$$\frac{\Delta\BR(\H\to\bbbar) }{\BR(\H\to\bbbar) }=2-3\%\mbox{~~~and~~~}
\frac{\Delta\BR(\H\to\gg) }{\BR(\H\to\gg) }=10-15\%$$
the total width of the Higgs boson can be calculated using
$$\Gamma_{\H}=\frac{[\Gamma(\H\to\gg) \BR (\H\to\bbbar)]}{
[\BR(\H\to\gg)] [\BR(\H\to\bbbar)]}$$ 
to an accuracy dominated by the expected
error on $\BR(\h\to\gamma\gamma)$.

The influence of the values of the b tag efficiencies for 
$\bbbar$ and $\ccbar$ events on
the accuracy of two-photon Higgs width was also studied.
b and c tag efficiencies were taken from 
a parametrisation by Battaglia \cite{Battaglia}.
In the region of b tag efficiencies from 50\% to 90\% the relative
error on $\Gamma(\h\to\gg)\BR(\h\to \bbbar)$ is quite stable.

\section{Conclusions}

Our preliminary results show that the two-photon width of the Higgs
boson can be measured at the photon-photon collider with high statistical 
accuracy of about 2-9\% for the Higgs mass range between 120 GeV and 160 GeV.
At such an accuracy one can discriminate between the SM
Higgs particle and the lightest scalar Higgs boson of the MSSM in the
decoupling limit, where all other Higgs bosons are very heavy and no
supersymmetric particle has been discovered at the e$^+$e$^-$ LC. Due to
the large charm production cross-section in $\gamma\gamma$ collisions,
excellent b tagging is required.  


\newpage

\begin{table*}[hbt]
\caption{Cross-sections for the Higgs signal and for the
background from direct heavy quark production in
$\gamma\gamma$ collisions. The $\gamma\gamma\to\bbbar (\g),\ccbar (\g)$
background was simulated for $\gamma\gamma$ invariant masses
larger than $W_{\rm min}$. A cut $\cos\theta<0.9$ was applied.
To calculate the event rates cross sections should be multiplied by the 
total luminosity 
$L_{\gamma\gamma}(0<z<z_{\max})$ defined in Section~3.}
\begin{tabular}{|c|c|c|c|c|c|c|}\hline
$m_{\h}$ & $\sqrt{s_{\rm ee}}$ & $\BR(\h\to\bbbar)$& $\sigma(\gamma\gamma\to\h\to \bbbar)$& 
$W_{\rm min}$ & $\sigma_{\rm \bar bb(g)}$ 
& $\sigma_{\rm \bar cc(g)}$ \\ 
(GeV)  & (GeV) & (\%) & (pb) & (GeV) & (pb) & (pb) \\ 
\hline
120 & 152.0 & 68  & 0.140 & \pz80 & 0.69 & 10.93 \\
\hline
140 & 177.0 & 34  & 0.089 & \pz90 & 0.57 & $\pz$8.87 \\
\hline
160 & 202.5 & 3.8 & 0.018 & 105   & 0.41 & $\pz$6.19 \\
\hline
\end{tabular}
\end{table*}

\newpage

\begin{figure*}[htb]
\begin{center}
\epsfig{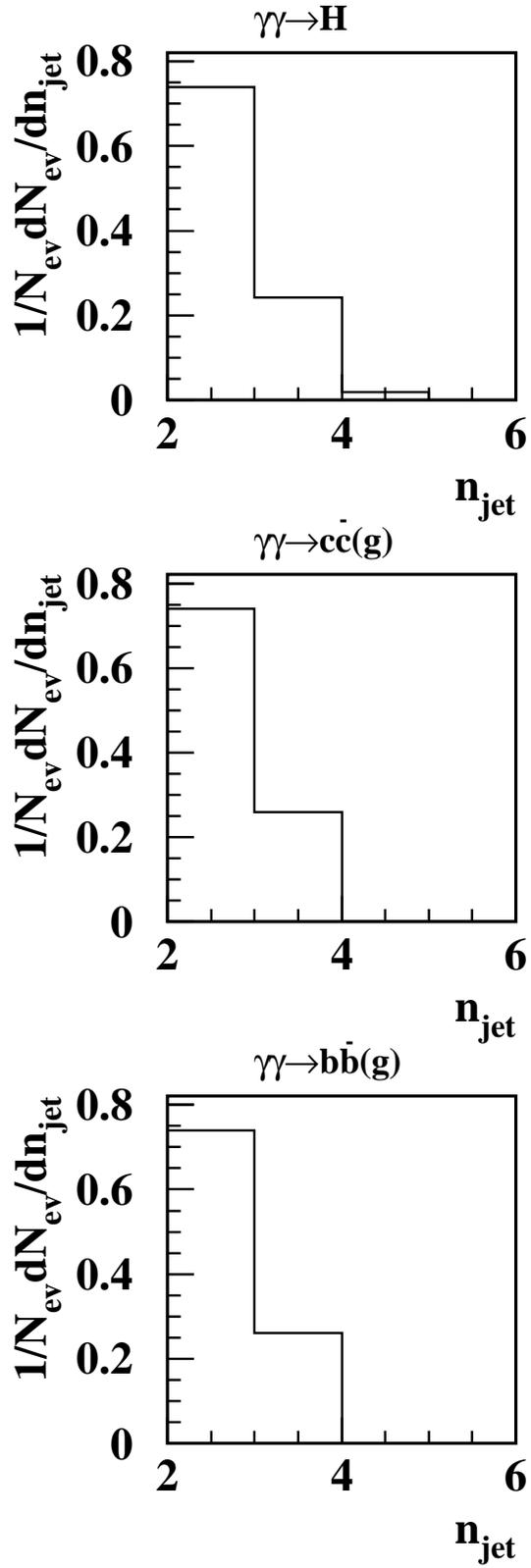}
\end{center}
\caption{Jet multiplicities for two- and three-parton event (detector level).}
\label{fig:multiplicities}
\end{figure*}

\newpage

\begin{figure*}[htb]
\begin{center}
\epsfig{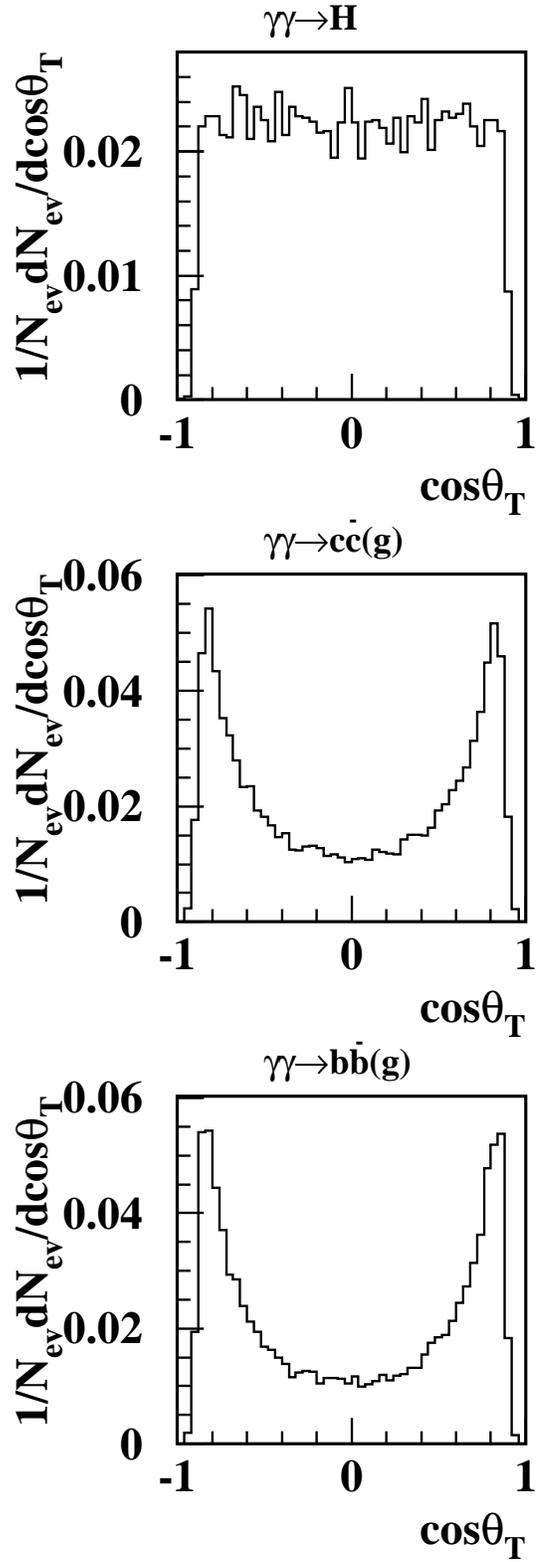}
\caption{Distribution of the thrust angle $\cos\theta_T$ for two- and 
three-parton events (detector level).}
\end{center}
\label{fig:cos}
\end{figure*}

\newpage

\begin{figure}[htb]
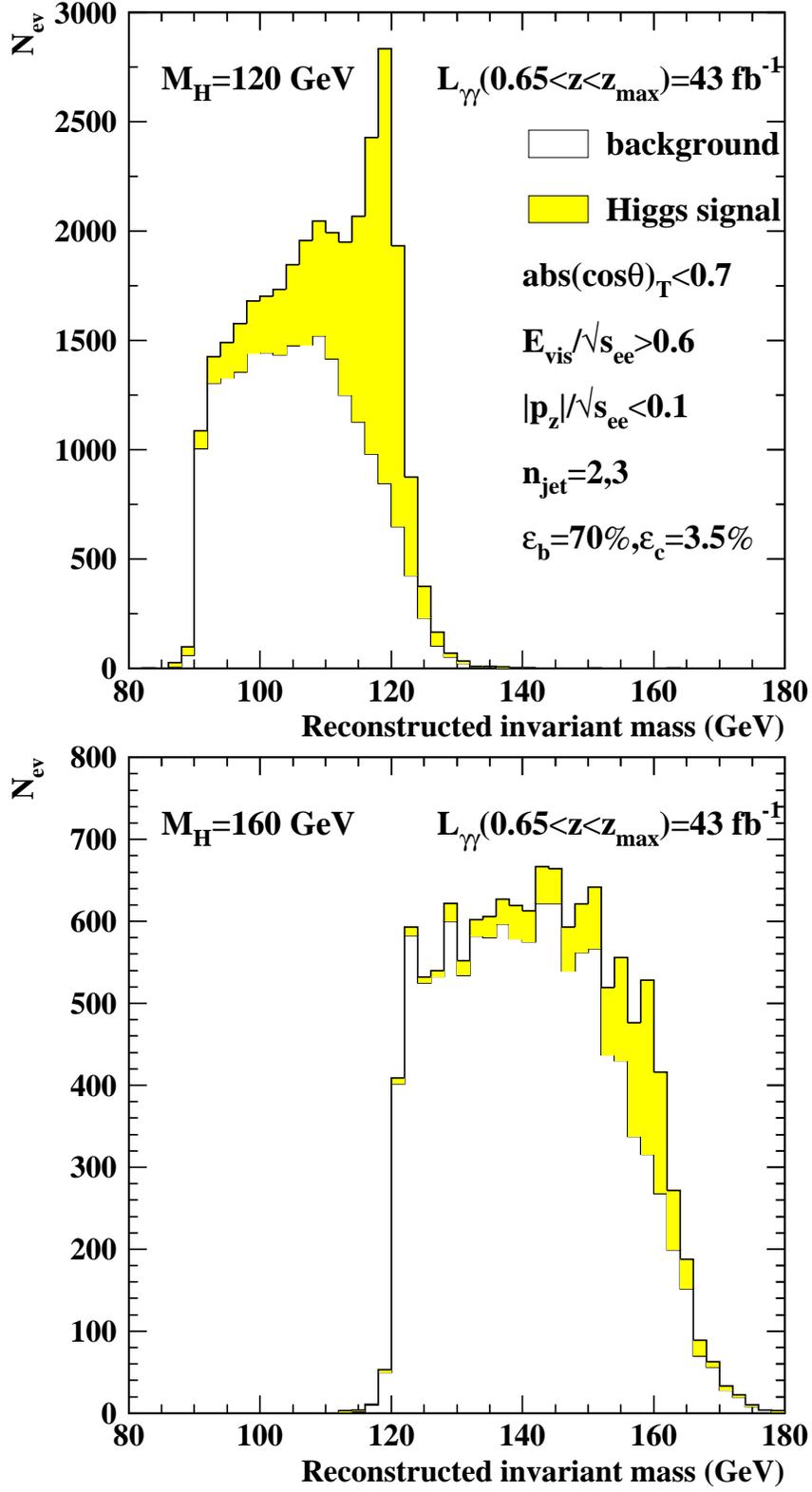

\begin{center}
\epsfig{file=fig5_120.epsi,width=0.8\textwidth}
\epsfig{file=fig5_160.epsi,width=0.8\textwidth}
\end{center}
\caption{Mass distributions for Higgs signal and heavy quark
background for a) $m_{\rm h}=120$~GeV and b) $m_{\rm h}=160$~GeV (detector level).} 
\label{fig:higgs}
\end{figure}


\begin{thebibliography}{99}
\bibitem{snow96}
J.F.~Gunion, L.~Poggioli, R.~Van~Kooten, C.~Kao, P.~Rowson, {\em
Proc. of the 1996 DPF/DPB Summer Study on ``New Directions in
High Energy Physics''}, June 25 - July 12, 1996,
Snowmass, Colorado, March 1997, UCD-97-5, hep-ph/9703330.
\bibitem{pik} 
P.~Igo-Kemenes, presentation given to
the LEP Experiments Committee open session, 
\mbox{\em http://lephiggs.web.cern.ch/LEPHIGGS/talks/index.html},
3~November 2000.
\bibitem{osaka}
B.~Pietrzyk, {\em The global fit to Electroweak data}, 
XXXth International Conference on High Energy Physics
27 July 27 -- 2 August 2000, Osaka, Japan,
http://www.ichep2000.rl.ac.uk/
\bibitem{MSSM}
S. Heinemeyer, W. Hollik, G. Weiglein, Eur. Phys. J. C9 (1999) 343; 
H E. Haber, to be published in the proceedings of 4th Int.
Symposium on Radiative Corrections, Barcelona, Spain, 8-12
September 1998, hep-ph/9901365.
\bibitem{melles}
M.~Melles, Nucl.~Phys.~B~(Proc.~Suppl.) 82 (2000) 379. 
\bibitem{watanabe}
I.~Watanabe et. al., KEK-REPORT-97-17, March 1998.
\bibitem{WW*} 
E.~Boos, V.~Ilyin, D.~Kovalenko, T.~Ohl, A.~Pukhov, M.~Sachwitz, 
H.J.~Schreiber, Phys. Lett. B427 (1998) 189.
\bibitem{ZZ} 
G.~Jikia, Phys.Lett. B298 (1993) 224;
Nucl. Phys.  B405 (1993) 24;
M.S.~Berger Phys. Rev.  D48 (1993) 5121;
D.A.~Dicus and C.~Kao, Phys. Rev.  D49 (1994) 1265.
\bibitem{decoupling} 
A. Djouadi, V. Driesen, W. Hollik, J.I. Illana,
Eur. Phys. J. C1 (1998) 149.
\bibitem{HDECAY} A.Djouadi, J.Kalinowski and M.Spira,
Comp.~Phys.~Comm.~108 (1998) 56.
\bibitem{BBC} D. Borden, D. Bauer, D. Caldwell, {\it SLAC-PUB}-5715,
{\it UCSD-HEP}-92-01.
\bibitem{JikiaTkabladze} G. Jikia, A. Tkabladze,
Proc. of the {\em Workshop on  gamma--gamma colliders}, March 28-31,
1994, Lawrence Berkeley Laboratory, Nucl. Instr. Meth.
A355 (1995) 81;
Phys. Rev.  D54 (1996) 2030;
\bibitem{Sudakov} 
M. Melles, W.J. Stirling, Phys. Rev. D59 (1999) 94009; 
Eur. Phys. J. C9 (1999) 101; 
M. Melles, W.J. Stirling, V.A. Khoze, hep-ph/9907238;
M. Melles, these proceedings, hep-ph/0008125.
\bibitem{PYTHIA} T.~Sj\"ostrand, Comp. Phys. Comm. 82 (1994) 74. 
\bibitem{BASES} S.~Kawabata, Comp. Phys. Comm. 88 (1995) 309.
\bibitem{luminosity} I.F.~Ginzburg, G.L.~Kotkin, S.L.~Panfil, V.G.~Serbo, 
V.I.~Telnov, Nucl. Instr. Meth. 219 (1984) 10.
\bibitem{Telnov}
V.I.~Telnov, Nucl.~Phys.~B~(Proc.~Suppl.) 82 (2000) 353. 
\bibitem{highlumi1}
V. Telnov, these proceedings, hep-ex/0010033
\bibitem{highlumi2}
W. Decking, these proceedings.
\bibitem{SIMDET} M.~Pohl, H.J.~Schreiber, 
DESY-99-030. 
\bibitem{Battaglia} 
M.~Battaglia, private communications.
\bibitem{hawkings}
R.~Hawkings, {\em Vertex detector and flavour tagging studies for the
TESLA linear collider},  LC-PHSM-2000-021

\end{thebibliography}
\end{document}